# Decomposition of Abraham momentum into complex orbital and spin momenta in evanescent wave

Jinbing Hu,[1] Songling Zhuang,[1] Hanming Guo[1,*]

[1]*Engineering Research Center of Optical Instrument and System, Ministry of Education; Shanghai Key Lab of Modern Optical System, College of Optical-Electrical Information and Computer Engineering, University of Shanghai for Science and Technology, Shanghai 200093, China*
*Corresponding author: hmguo@usst.edu.cn*

The Abraham momentum of electromagnetic field represents the decomposition property into real-valued orbital and spin contributions. However, we find that the orbital and spin momenta of evanescent electromagnetic fields are inherently complex. Consequently, the decomposition of Abraham momentum is reexamined and the exact expressions of orbital and spin momenta are obtained, which are applicable to both propagating and evanescent electromagnetic fields. Furthermore, we also justify the newly-derived decomposition of Abraham momentum in lossy media, which is further demonstrated by a concrete example of surface polariton of metal. In addition, on the basis of the complex orbital momentum of evanescent fields the complex group velocity is tentatively put forward with its real part representing the energy flux velocity and its imaginary part denoting the osmotic velocity of energy penetrating the corresponding media.

Recently the decomposition of Abraham momentum of electromagnetic fields into orbital and spin parts attracts enormous attentions [1-5]. On one hand, the orbital and spin contributions of Abraham momentum, respectively, produce orbital and spin angular momentum (AM)[5-7], which play distinct roles in the interaction with matters, like atoms[8-10] and small particles[11-15]. For instance, the orbital AM of light beam makes the particle rotate around the axis of beam[11, 12], and one can operate single atom by transferring the spin AM of photon to the atom within a high-finesse cavity[8-10]. These effects basically underlie many applications, such as quantum computation[16, 17] and quantum information process[18]. On the other hand, it has been experimentally demonstrated that it is the orbital momentum rather than the Abraham momentum that represents the physically-meaningful momentum density of light beam (only in linearly polarized paraxial fields and plane waves is the orbital momentum equivalent to Abraham momentum)[13, 19]. For instance, the orbital momentum of light beam exerts optical forces on small particles immerged in this light beam[20]. On basis of this effect, many practical applications can be achieved, such as laser cooling[21-23], optical manipulation of atoms or small particles[24-27].

The decomposition of the Poynting vector (Abraham momentum) into orbital and spin contributions for paraxial fields was first put forward by Bekshaev and Soskin[1]. Then, M. V. Berry[2] generalized in 2009 this decomposition to monochromatically nonparaxial fields and obtained the following expressions for orbital and spin parts in terms of electric-magnetic decocracy, , i.e. $\mathbf{P} = \mathbf{P}^o + \mathbf{P}^s$ with:

$$\mathbf{P}^o = (g/2\omega)\operatorname{Im}[\mu^{-1}\mathbf{E}^* \cdot (\nabla)\mathbf{E} + \varepsilon^{-1}\mathbf{H}^* \cdot (\nabla)\mathbf{H}], \quad (1)$$

$$\mathbf{P}^s = (g/4\omega)\nabla \times \operatorname{Im}[\mu^{-1}\mathbf{E}^* \times \mathbf{E} + \varepsilon^{-1}\mathbf{H}^* \times \mathbf{H}]. \quad (2)$$

Here, $g = 1/8\pi$, $\omega$ is the angular frequency, and $\varepsilon$ $\mu$ are, respectively, the permittivity and permeability of medium where electromagnetic field propagates. $\mathbf{E}(\mathbf{r})$ and $\mathbf{H}(\mathbf{r})$ are, respectively, the total full-time-independent three-dimensional complex electric and magnetic fields, including both the excited and polarized fields.

In the same year, unlike the differential method Berry used, Li[5] also obtained the same decomposition of Abraham momentum into orbital and spin parts by expressing the electric and magnetic field in free space as integrals over the plane-wave spectrum in reciprocal space. Since then, the decomposition of momentum of light attracts numerous attentions [1-4]. As explicitly pointed out the above division can only be applicable to propagating fields[2, 5]. However, K. Y. Bliokh et al directly extended this decomposition to the case of evanescent fields and obtained purely real-valued orbital momentum in evanescent field[3, 4]. For instance they achieved an orbital momentum $\mathbf{P}^o = (w/\omega)k_z\hat{\mathbf{z}}$ in the evanescent wave with complex wavevector $\mathbf{k} = k_z\hat{\mathbf{z}} + i\kappa\hat{\mathbf{x}}$, where $w$ is the spatially-inhomogeneous energy density of evanescent wave[3]. As we know from the field theory, for vectorial field (not in intensity singularities) there is such a relationship between orbital momentum and wavevector[2, 13]

$$\mathbf{k} = \mathbf{P}^o / \langle \vec{\psi} | \vec{\psi} \rangle. \quad (3)$$

Here, $\vec{\psi}$ is the vectorial field function. Therefore, the orbital momentum of evanescent fields should be complex as well since its wavevector is complex. Now the question arises, how does this difference happen.

Due to the practically significant applications of orbital and spin momenta it is of great importance to resolve above question. In the present letter, starting from the definition of Abraham momentum of monochromatically electromagnetic field, we will first reexamine the decomposition of Abraham momentum and derive more exact expressions for its orbital and spin contributions. Then, by comparing the newly-derived expressions of orbital and spin momenta with previous ones i.e. Eqs. (1) and (2), we will give the physical origin why the previous expressions are not applicable to the cases of evanescent electromagnetic field, such as surface plasmon polaritons (SPPs) of metal, evanescent wave arising from total internal reflection and so on. Furthermore, concerning the statement by R. W. Ziolkowski *et al*[28] that the decomposition of kinetic (Abraham) momentum of electromagnetic fields in lossy material is of no physical meaning, we will demonstrate that the decomposition of Abraham momentum into orbital and spin parts is applicable not only to lossless media, but also to lossy material, like single complex materials with $\varepsilon = \varepsilon_r + i\varepsilon_i (\varepsilon_r < 0)$ or $\mu = \mu_r + i\mu_i (\mu_r < 0)$; then, a transverse magnetic (TM) surface wave at the interface between a dielectric medium and a metal is taken as a demonstration. The last but not the least, according to the newly-derived orbital momentum of evanescent field the complex group velocity is tentatively put forward, of which the imaginary part turn out to have the meaning of osmotic velocity as extensively studied in field theory.

First of all, we reexamine the decomposition of Abraham momentum of monochromatically electromagnetic field subject to the following Maxwell equations in a uniform nondispersive medium with permittivity $\varepsilon$ and permeability $\mu$

$$-i\omega\varepsilon\mathbf{E} = c\nabla\times\mathbf{H} \quad \nabla\cdot\mathbf{E} = 0 \quad . \quad (4)$$
$$-i\omega\mu\mathbf{H} = -c\nabla\times\mathbf{E} \quad \nabla\cdot\mathbf{H} = 0$$

Here, $c$ is the speed of light in vacuum, and we use Gaussian units. The time-averaged Abraham momentum density of above monochromatic field is known to be[29]

$$\mathbf{P} = (g/c)\mathrm{Re}[\mathbf{E}^*\times\mathbf{H}]. \quad (5)$$

To avoid any loss of physically-meaningful information we guarantee each step being identity transformation in following procedure. Eliminating the Re-operation in Eq. (5) in terms of the identity $2\mathrm{Re}[\mathbf{A}] = \mathbf{A}^* + \mathbf{A}$ ( $\mathbf{A} = \mathbf{E}^*\times\mathbf{H}$ ), the time-averaged momentum density can be written equivalently as

$$\mathbf{P} = (g/2c)\left[\mathbf{E}^*\times\mathbf{H} - \mathbf{H}^*\times\mathbf{E}\right]. \quad (6)$$

Replacing the magnetic (electric) field in Eq. (6) with its corresponding curl equation of electric (magnetic) field, Eq. (6) reads

$$\mathbf{P} = (g/2i\omega)[\mu^{-1}\mathbf{E}^*\times(\nabla\times\mathbf{E}) + \varepsilon^{-1}\mathbf{H}^*\times(\nabla\times\mathbf{H})]. \quad (7)$$

In a further step, making use of the vector calculus identity $\mathbf{A}^*\times(\nabla\times\mathbf{A}) = \mathbf{A}^*\cdot(\nabla)\mathbf{A} - (\mathbf{A}^*\cdot\nabla)\mathbf{A}$ ( $\mathbf{A} = \mathbf{E},\mathbf{H}$ ), which can be derived as a linear combination of commonly used vector calculi[30] although it is not commonly used, the Abraham momentum Eq. (7) can be further written as

$$\mathbf{P} = (g/2i\omega)\left[\mu^{-1}\mathbf{E}^*\cdot(\nabla)\mathbf{E} + \varepsilon^{-1}\mathbf{H}^*\cdot(\nabla)\mathbf{H}\right]$$
$$+ (gi/2\omega)\left[\mu^{-1}(\mathbf{E}^*\cdot\nabla)\mathbf{E} + \varepsilon^{-1}(\mathbf{H}^*\cdot\nabla)\mathbf{H}\right]. \quad (8)$$

According to the field theory, the momentum of vectorial field can be decomposed into its orbital and spin contributions, producing their respective angular momentum[31]. So, apparently the orbital and spin momentum densities of the monochromatically electromagnetic field can be respectively written as

$$\tilde{\mathbf{P}}^o = (g/2i\omega)\left[\mu^{-1}\mathbf{E}^*\cdot(\nabla)\mathbf{E} + \varepsilon^{-1}\mathbf{H}^*\cdot(\nabla)\mathbf{H}\right], \quad (9)$$
$$\tilde{\mathbf{P}}^s = (ig/2\omega)\left[\mu^{-1}(\mathbf{E}^*\cdot\nabla)\mathbf{E} + \varepsilon^{-1}(\mathbf{H}^*\cdot\nabla)\mathbf{H}\right]. \quad (10)$$

With careful observation we can see the only difference between Eqs. (9) and (1) is the replacing of Im-operation in Eq. (1) by "$1/i$" in Eq. (9). This difference seems to be negligible, but actually it can result in physically different consequence, especially for evanescent fields; because in evanescent fields Eq. (9) yields complex orbital momentum while Eq. (1) predicts a purely real-valued one. The difference can be even more apparent if we write down the real and imaginary parts of Eq. (9)

$$\mathrm{Re}\left[\tilde{\mathbf{P}}^o\right] = (g/2\omega)\mathrm{Im}\left[\mu^{-1}\mathbf{E}^*\cdot(\nabla)\mathbf{E} + \varepsilon^{-1}\mathbf{H}^*\cdot(\nabla)\mathbf{H}\right], \quad (11a)$$
$$\mathrm{Im}\left[\tilde{\mathbf{P}}^o\right] = -(g/2\omega)\mathrm{Re}\left[\mu^{-1}\mathbf{E}^*\cdot(\nabla)\mathbf{E} + \varepsilon^{-1}\mathbf{H}^*\cdot(\nabla)\mathbf{H}\right]. \quad (11b)$$

We can clearly see that, apart from the real part which is exact same as Eq. (1), the newly-derived orbital momentum in Eq. (9) includes an imaginary component (i.e. Eq. (11b)) as well, which is the exact quantity omitted in the treatment of evanescent wave by K. Y. Bliohk *et al*, as will be demonstrated later.

The main reason resulting in this difference is because of the presence of the vector differential operator $\nabla$ in the quantity $\mathbf{A}^*\cdot(\nabla)\mathbf{A}$ ( $\mathbf{A} = \mathbf{E},\mathbf{H}$ ), the result of which, besides the intensity, includes a product of "$i$" and the wavevector $\mathbf{k}$. Then, the Im-operation in Eqs. (1) and (2) keeps the imaginary part of this quantity. We know the Im-operation in Eqs. (1) and (2) originates from the Re-operation in the definition of Abraham momentum, which is taken to ensure the practically physical meaning of results obtained when the electric and magnetic fields are expressed in exponential form. This is true for linear operation of electromagnetic field, but not true for the situations in Eqs. (1) and (2) due to the presence of the vector differential operator $\nabla$. Provided the wavevector is complex, e.g., $\mathbf{k} = k_z\hat{\mathbf{z}} + i\kappa\hat{\mathbf{x}}$, both its real and imaginary parts are of physical meaning, and of which neither the real nor the imaginary should be dumped. In a word, for propagating electromagnetic field there is no problem in performing such Im-operation and Eqs. (1) and (2) yield correct results of orbital and spin momenta; while for evanescent field such Im-operation keeps only the imaginary part and dumps the real part of the quantity in the square bracket, i.e. throwing the imaginary part of orbital and spin momenta. Therefore, the decomposition of the Abraham momentum into orbital and spin parts liking the manner of Eqs. (1) and (2) is not physically rigorous; some important components are omitted in the transformation from Re-operation to Im-operation.

To further demonstrate above conclusion and for better comprehension, we take into account the same electromagnetic wave (i.e. Eq. (7) in Ref. 3), a single evanescent wave propagating along the z-axis and decaying in the $x > 0$ half space. But we differentiate the general imaginary unit "$i$" from that in complex wavevector, e.g. $\mathbf{k} = k_z\hat{\mathbf{z}} + j\kappa\hat{\mathbf{x}}$, then the electric field and its corresponding magnetic field take the following form

$$\mathbf{E} = \frac{A\sqrt{\mu}}{\sqrt{1+|m|^2}}\left(\hat{\mathbf{x}} + m\frac{k}{k_z}\hat{\mathbf{y}} - j\frac{\kappa}{k_z}\hat{\mathbf{z}}\right)\exp(ik_zz + ij\kappa x), \quad (12)$$

$$\mathbf{H} = \frac{A\sqrt{\varepsilon}}{\sqrt{1+|m|^2}}\left(\frac{k}{k_z}\hat{\mathbf{y}} - m\hat{\mathbf{x}} + mj\frac{\kappa}{k_z}\hat{\mathbf{z}}\right)\exp(ik_zz + ij\kappa x). \quad (13)$$

Here, $A$ is the wave amplitude, $k_z = k\cosh\vartheta > k$ is the longitudinal wave number, and $\kappa = k\sinh\vartheta$ is the exponential decay rate. Substitute Eqs. (12) and (13) into Eq. (1), we reach

$$\mu^{-1}\mathbf{E}^* \cdot (\nabla)\mathbf{E} + \varepsilon^{-1}\mathbf{H}^* \cdot (\nabla)\mathbf{H} = 2A^2 i(k_z\hat{\mathbf{z}} + j\kappa\hat{\mathbf{x}})\exp(2ij\kappa x), \quad (14)$$

which clearly shows the product of "$i$" and the complex wavevector $\mathbf{k} = k_z\hat{\mathbf{z}} + j\kappa\hat{\mathbf{x}}$. Here we are in the critical point. As mentioned before, both the real and imaginary parts of complex wavevector are of phycial meaning, neither should be dumped; therefore, the Im-operation in Eq. (1) is only valid to general imaginary unit "$i$". Under this condition, we have the orbital momentum of the evanescent field

$$\tilde{\mathbf{P}}^o = (w/\omega)(k_z\hat{\mathbf{z}} + j\kappa\hat{\mathbf{x}}). \quad (15)$$

As expected, the obtained orbital momentum is now proportional to the complex wavevector, in agreement with field theory.

On the other hand, if we substitute Eqs. (12) and (13) into Eq. (9) we can directly obtain the same result as Eq. (15) without the above restriction on imaginary unit, revealing that Eq. (9) is the exact expression of orbital momentum density. In fact, from the derivation procedure we can conclude that both Eqs. (9) and (10) are, respectively, the exact expressions of orbital and spin momenta as no restriction is imposed to the electric and magnetic fields.

Through above analysis we can conclude that to obtain physically-meaningful quantity Re- or Im-operation should be carefully taken when there is the vector differential operator $\nabla$ operating on electromagnetic field because the vector differential operator has both differential and vectorial properties, the combination of which gives rise to the product of "$i$" and wavevector of fields. One useful suggestion regarding this point is to eliminate Re- or Im-operation using the identity $2\text{Re}[\mathbf{A}] = \mathbf{A} + \mathbf{A}^*$ or $2\text{Im}[\mathbf{A}] = -i(\mathbf{A} - \mathbf{A}^*)$.

Another important issue about the decomposition of Abraham momentum is whether this decomposition still holds true in lossy materials. In 2013, R. W. Ziolkowski et al [28] stated in their paper that "to the best knowledge of the authors, there has not been a decomposition of the kinetic momentum orbital and spin parts within a lossy medium reported". Their main reason is that the decomposition of Abraham momentum into orbital and spin parts in accordance to the vector fields (i.e. A(11) ~ A(14) in Ref. 28) losses its physical meaning for lossy media. According to their derivation procedure we can see that their decomposition of Abraham momentum is based on the prerequisite: the electric and magnetic polarizations are zero, i.e. $\mathcal{P}, \mathcal{M} = 0$. Therefore, in accordance to their method, such decomposition only holds true in vacuum or free space. For electrically or/and magnetically polarizable media, the polarizations $\mathcal{P}, \mathcal{M}$ are not zero, so the decomposition losses the validity, let along in lossy media. The main reason for this paradox is that they took the electric and magnetic fields in Abraham momentum definition as the excited fields rather than the total fields.

Actually, the decomposition of Abraham momentum into orbital and spin contributions is applicable not only to lossless media, but also to lossy media, say single complex materials with $\varepsilon = \varepsilon_r + i\varepsilon_i(\varepsilon_r < 0)$ or $\mu = \mu_r + i\mu_i(\mu_r < 0)$. For electromagnetic fields in lossless media, the wavevector is a real-valued quantity, and the quantities in square brackets in Eqs. (9) and (10) are purely imaginary. Eq. (9) is then equivalent to Eq. (1). Meanwhile, Eq. (10), associated with the vector calculus identity $\text{Im}\left[(\mathbf{A}^* \cdot \nabla)\mathbf{A}\right] = (1/2i)\nabla \times (\mathbf{A} \times \mathbf{A}^*)$, is also physically equivalent to Eq. (2). Therefore, such decomposition in lossless media remains its validity.

As for the case of electromagnetic fields in single complex media, the rewrite of complex orbital momentum Eq. (9) as the sum of its real and imaginary parts is still valid. However, the relationship between Eqs. (10) and (2) is not so straightforward, but reverse thinking will be helpful for understanding this. Assuming that the permittivity and permeability are real, Eq. (10) can also be written as the sum of its real and imaginary parts $\tilde{\mathbf{P}}^s = \text{Re}[\tilde{\mathbf{P}}^s] + \text{Im}[\tilde{\mathbf{P}}^s]$ with

$$\text{Re}\left[\tilde{\mathbf{P}}^s\right] = (g/4\omega)\nabla \times \text{Im}\left[\mu^{-1}(\mathbf{E}^* \times \mathbf{E}) + \varepsilon^{-1}(\mathbf{H}^* \times \mathbf{H})\right], \quad (16a)$$

$$\text{Im}\left[\tilde{\mathbf{P}}^s\right] = (g/2\omega)\text{Re}\left[\mu^{-1}(\mathbf{E}^* \cdot \nabla)\mathbf{E} + \varepsilon^{-1}(\mathbf{H}^* \cdot \nabla)\mathbf{H}\right]. \quad (16b)$$

Then, we consider the structures consisting of single complex medium and lossless medium; more precisely, the structures can be divided into two cases: one composed by media with $\varepsilon_1 = \varepsilon_r + i\varepsilon_i, \varepsilon_r < 0$ and medium $\varepsilon_2 > 0$, $\mu_1 = \mu_2 = 1$; the other by media with $\mu_1 = \mu_r + i\mu_i, \mu_r < 0$ and medium with $\mu_2 > 0$, $\varepsilon_1, \varepsilon_2 > 0$. The above structures support only transverse-magnetic (TM) and transverse electric (TE) surface modes respectively. Without any loss of generality, we consider TM surface mode with the electric field being complex and the magnetic field being purely real, then the cross product of magnetic field yields zero, i.e. $\mathbf{H}^* \times \mathbf{H} = 0$. That is, the second term in Eq. (16a) is zero. Now that $\mu_1 = \mu_2 = 1$ being real-valued quantities, there is $\text{Im}[\mu^{-1}(\mathbf{E}^* \cdot \nabla)\mathbf{E}] = (1/2i\mu)\nabla \times (\mathbf{E}^* \times \mathbf{E})$. Namely, even in single complex material sustaining of TM surface wave the real part of Eq. (10), i.e. Eq. (16a) gives the physical meaning of spin momentum density. Similar analysis can be done to single complex material sustaining of TE surface wave. In a word, no matter in which case it is, Eq. (10) can be written as the sum of its real and imaginary parts with the real part representing spin momentum density. Therefore, the decomposition of Abraham momentum into orbital and spin parts is still valid in single complex media.

To further prove above conclusion, we take TM surface plasmon polariton (SPP) at the $x = 0$ interface between the dielectric medium with permittivity $\varepsilon_+ > 0$ ($x > 0$) and a metal ($x < 0$) with permittivity $\varepsilon_-$, along with $\mu_+ = \mu_- = 1$ as an example. Assuming that the surface wave propagates along the $z$-axis, then its electric and magnetic fields can be written as[32]

$$\begin{aligned}\mathbf{E}_\pm &= \varepsilon_\mp (k_z\hat{\mathbf{x}} \mp i\kappa_\pm\hat{\mathbf{z}})\exp(ik_z z \mp \kappa_\pm x) \\ \mathbf{H}_\pm &= \varepsilon_+\varepsilon_- k_0 \exp(ik_z z \mp \kappa_\pm x)\hat{\mathbf{y}}\end{aligned}. \quad (17)$$

Here, the "+" and "−" subscripts denote quantities in $x > 0$ and $x < 0$ half spaces. The surface waves are characterized by complex wave vectors $\mathbf{k}_\pm = k_z\hat{\mathbf{z}} \pm i\kappa_\pm\hat{\mathbf{x}}$, which satisfy the dispersion relations $k_\pm^2 = k_z^2 - \kappa_\pm^2$. In accordance to the boundary condition of electromagnetic wave, there is the following relationship for SPP parameters[32]

$$k_z = k_0\sqrt{\varepsilon_+\varepsilon_-/(\varepsilon_+ + \varepsilon_-)} \text{, and } \kappa_+\varepsilon_- = -\varepsilon_+\kappa_- . \quad (18)$$

Here $k_0 = \omega_0/c$ is the wave number in vacuum and $\omega_0$ is the angular frequency. It should be stressed that the electric field expressed in Eqs. (12) and (13) describe not SPP, but evanescent wave generated through total internal reflection of linearly polarized plane wave at a desired polarization angle (see Supplementary of Ref. 3 for more detail).

Substituting the electric and magnetic fields of Eqs. (17) into Eq. (9) and taking some simple operation yields the orbital momentum density of the SPP

$$\tilde{\mathbf{P}}^o_\pm = (g/\omega)\varepsilon_\mp^2 k_z^2 \exp(\mp 2\kappa_\pm x)(k_z\hat{\mathbf{z}} \pm i\kappa_\pm\hat{\mathbf{x}}). \quad (19)$$

It is clear that Eq. (19) still remain the proportionality between orbital momentum and wavevector of SPP although they are both complex, further implying the generality of field theory. Similarly, we can get the time-averaged Abraham and spin momentum densities of such SPP

$$\mathbf{P}_\pm = (g/\omega)\varepsilon_\mp^2\varepsilon_\pm k_z k_0^2 \exp(\mp 2\kappa_\pm x)\hat{\mathbf{z}}, \quad (20)$$

$$\tilde{\mathbf{P}}^s_\pm = (g/\omega)\varepsilon_\mp^2 k_z\kappa_\pm \exp(\mp 2\kappa_\pm x)(\mp i k_z\hat{\mathbf{x}} - \kappa_\pm\hat{\mathbf{z}}). \quad (21)$$

From Eqs. (19) ~ (21) we can easily identify the decomposition of Abraham momentum into orbital and spin contributions in the case of SPP propagating in metal, a lossy medium. More precisely, the x components of orbital and spin momenta are of the same value but with opposite sign, the difference in z components of orbital and spin momenta gives rise to the Abraham momentum of surface wave.

For better understanding, we plot in Fig. 1 the momenta flux of specific SPP propagating at the surface of metal ($x<0$) with permittivity $\varepsilon_- = -1.5 + 0.1i$, the $x>0$ half space is vacuum. In vacuum the z component of orbital momentum is larger than that of spin momentum, resulting in positive-z-direction Abraham momentum; while in metal the situation is opposite, resulting in negative-z-direction Abraham momentum, see left figure of Fig. 1. Globally, the total momentum flux forms a circulation around the interface, as shown in right figure of Fig. 1, which is in agreement with the conclusion of Lai[33].

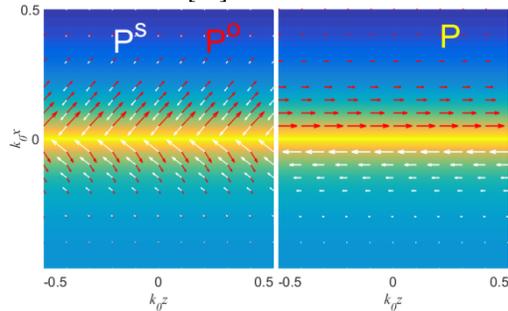

Fig. 1 (Left) the orbital (red) and spin (white) momentum flux of specific SPP propagating along the surface of metal ($x<0$) with $\varepsilon_- = -1.5 + 0.1i$, the $x>0$ half space is vacuum. (Right) the total (i.e. Abraham) momentum of the specific SPP. Notice the arrows in right figure are enlarged by 4 times.

In addition, we notice another important feature that the group velocity of SPP is complex as well. It has now been theoretically[3] and experimentally[13] demonstrated that it is the orbital momentum $\tilde{\mathbf{P}}^o$ rather than the Poynting vector that represents the physically-meaningful momentum density of light; and divided by the energy density w, the orbital momentum yields local group velocity. So the local group velocity of SPP can be written as $\tilde{\mathbf{v}}^g_\pm = \tilde{\mathbf{P}}^o c^2/w$. Taking the complex orbital momentum Eq. (19) into this expression, we have

$$\tilde{\mathbf{v}}^g_\pm = (c^2/\omega\varepsilon_\pm)(k_z\hat{\mathbf{z}} \pm i\kappa_\pm\hat{\mathbf{x}}). \quad (22)$$

Apparently, the group velocity of SPP is also proportional to the complex wavevector. As we have known the z-component velocity represents the energy flux along the interface; as for the x-component velocity arising from the imaginary part of complex orbital momentum, it is nothing but the osmotic velocity, representing the velocity of energy penetrating into corresponding medium, which has been extensively discussed in field theory [34-37]. However, we would like to point out that the energy in vacuum side flows in +z direction with superluminal velocity $|\tilde{\mathbf{v}}^g_+| = ck_z/k_0 > c$ w[3, 4] and osmotic velocity is along +x direction, while in metal side due to the negative feature of permittivity the energy flux and osmotic velocities are in opposite direction as that in vacuum.

In conclusion, from the definition of Abraham momentum, we reexamined the decomposition of Abraham momentum of monochromatically electromagnetic fields, and gave the exact expressions for orbital and spin momenta that are applicable to both propagating and evanescent fields. Then, the exactly physical reason why the previous decomposition of Abraham momentum into orbital and spin parts does not hold true for evanescent fields was given. Based on the exact expressions we demonstrated that the decomposition of Abraham momentum into orbital and spin contributions is valid in both lossless and lossy media. In addition, we revealed that the complex orbital momentum of SPP gives rise to complex local group velocity, the imaginary part of which represents osmotic velocity. Finally, we note that the validity of the decomposition of Abraham momentum in another kind lossy media, such as with permittivity $\varepsilon = \varepsilon_1 + i\varepsilon_2, \varepsilon_1 > 0$, is much more intricate and due to the limitation of length we will report elsewhere.

Acknowledgement. We acknowledge support from National key research and development program of China (2016YFF0101603); National Basic Research Program of China (2015CB352001); National Natural Science Foundation of China (61178079); Leading Academic Discipline Project of Shanghai Municipal Government (S30502); Development program of University of Shanghai for Science and Technology.